\documentclass[preprint,showpacs,preprintnumbers,amssymb,superscriptaddress,aps,prd,nofootinbib,11pt]{revtex4-1}

\selectfont 

\usepackage{graphicx}       % Include figure files
\usepackage{dcolumn}        % Align table columns on decimal point
\usepackage{bm}             % bold math
\usepackage{psfrag}
\usepackage{tensor}
\usepackage[usenames]{color}
\definecolor{navyblue}{rgb}{0.0, 0.0, 0.5}
\usepackage[linktocpage,colorlinks=true,allcolors=navyblue]{hyperref}
\usepackage{amsmath}
\usepackage{caption}
\usepackage{subcaption}
\usepackage{physics}

\newcommand{\cD}{\mathcal{D}}

\newcommand{\cL}{\mathcal{L}}

\newcommand{\beq}{\begin{equation}}
\newcommand{\eeq}{\end{equation}}

\newcommand{\asm}{\mathrm{a}}   % We are using 'a' for the spin parameter and the semi-major axis. I have replaced the latter with \asm.
\newcommand{\rmin}{r_{\text{min}}}
\newcommand{\rmax}{r_{\text{max}}}

\begin{document}

\preprint{}

 \title{Adiabatic inspirals under electromagnetic radiation reaction on Kerr spacetime}

\author{Ethan J. German}
% \email{ejgerman1@sheffield.ac.uk}
\affiliation{Consortium for Fundamental Physics, School of Mathematics and Statistics,
University of Sheffield, Hicks Building, Hounsfield Road, Sheffield S3 7RH, United Kingdom}

\author{Kevin Cunningham}
% \email{kevin.cunningham1@ucdconnect.ie}
\affiliation{
School of Mathematical Sciences, University College Dublin, Belfield, Dublin 4, Ireland
}

\author{Visakan Balakumar}
% \email{v.balakumar@sheffield.ac.uk}
\affiliation{Consortium for Fundamental Physics, School of Mathematics and Statistics,
University of Sheffield, Hicks Building, Hounsfield Road, Sheffield S3 7RH, United Kingdom}

\author{Niels Warburton}
% \email{niels.warburton@ucd.ie}
\affiliation{
School of Mathematical Sciences, University College Dublin, Belfield, Dublin 4, Ireland
}

\author{Sam R. Dolan}
% \email{s.dolan@sheffield.ac.uk}
\affiliation{Consortium for Fundamental Physics, School of Mathematics and Statistics,
University of Sheffield, Hicks Building, Hounsfield Road, Sheffield S3 7RH, United Kingdom}

\date{\today}

\begin{abstract}
A compact body in orbit about a black hole loses orbital energy and angular momentum through radiation-reaction processes, inspiralling towards the black hole until a final plunge. Here we consider a scenario with a charged compact body in which fluxes of electromagnetic radiation drive this inspiral. We calculate trajectories in the $(p,e)$ plane for inspirals in the equatorial plane of a rotating black hole within the adiabatic (orbital-averaged-dissipative) approximation. We make comparisons with a non-relativistic Keplerian approximation based on the Abraham-Lorentz force law, and with standard gravitational-wave driven scenarios. We find that EM-driven inspirals are less efficiently circularized (i.e.~orbits remain more eccentric at the point of plunge) than their gravitational counterparts, and we quantify the effect of black hole spin.
\end{abstract}

\maketitle

\vspace{0.3cm}

\section{Introduction\label{sec:introduction}}

Since the celebrated first-light detection of 2015 \cite{LIGOScientific:2016aoc}, networks of gravitational-wave interferometers have observed numerous ``chirp'' signals originating from compact binaries during the final stages of inspiral and coalescence 
\cite{LIGOScientific:2016wyt, LIGOScientific:2021djp}. The emission of gravitational waves, which is strongest at periapsis, removes eccentricity from a binary system, causing the orbits to become more circular as the inspiral progresses \cite{Peters:1963ux,Peters:1964zz}. For comparable-mass binaries -- those accessible in the frequency band of ground-based detectors -- the circularisation process is efficient, and by the point of detection the bodies are on nearly-circular orbits. Thus, the observed signals are regular ``chirps'' characterised by a small number of parameters (chiefly, the chirp mass) \cite{Cutler:1994ys}. Conversely, extreme mass-ratio inspirals (EMRIs) are less efficiently circularised, and the orbit of the secondary is expected to remain significantly eccentric and non-equatorial all the way up to the plunge \cite{Gair:2017ynp}. EMRIs are key targets for space-based detectors (i.e.~LISA), which can probe much low frequencies due to the absence of seismic noise in space.

One approach to modelling EMRIs is by calculating the so-called self-force \cite{Poisson:2011nh,Pound:2021qin,Wardell:2015kea}. The essential idea is that the emission of gravitational waves is associated with a back-reaction on the motion of the system that is described by a force acting on the secondary (of mass $m$) moving on the spacetime of the primary (typically, a black hole of mass $M$ and spin $a = J/M$). A complementary view is that the motion is geodesic in a certain \emph{effective} spacetime \cite{Detweiler:2002mi}.
The self-force programme is well-developed at leading order in $m/M$ \cite{Pound:2021qin,Katz:2021yft}, with first results at second order in the mass ratio now available \cite{Pound:2019lzj, Warburton:2021kwk, Wardell:2021fyy}.

Self-force, as a concept, has historical roots in electromagnetism: it builds upon Dirac's idea of separating the field into its symmetric/singular and radiative/regular parts, with the latter generating the self-force \cite{Dirac:1938nz, Detweiler:2002mi, Poisson:2011nh}. Since 1997, the focus of the programme has been on calculating \emph{gravitational} self force (GSF), because this is the predominant driver of astrophysical binaries. However, the electromagnetic self force (ESF) and radiation reaction (associated with the dissipative part of ESF) is interesting in its own right, and has received some attention in the literature of recent years \cite{DeWitt:1964, Leaute:1982, Piazzese:1991, Gralla:2009md, Haas:2011np, Nolan-thesis, Torres:2020fye, Zhang:2023vok}. In this work, we seek to extend the ESF frontier by studying the evolution of the eccentric, equatorial orbit of a charged particle around a Kerr black hole for the first time (see Ref.~\cite{Torres:2020fye} the circular orbit case). 

There are at least two good reasons for considering the electromagnetic self force in binary systems. Firstly, it provides an opportunity to compare and contrast an electromagnetically-driven scenario with the standard gravitationally-driven scenario. A key difference is that electromagnetically-driven systems radiate principally in the dipole ($\ell=1$) mode, whereas gravitationally-driven systems radiate principally in the quadrupole ($\ell = 2$) mode. A consequence is that the rate of the increase of the orbital frequency during the inspiral, $f(t) \propto (-t)^{-\beta}$ (for $r \gg M$), is markedly different, with an index $\beta = 3/8$ in the gravitational case and an index $\beta = 1/2$ in the electromagnetic case \cite{LIGOScientific:2016wyt, Torres:2020fye}. A second consequence is that the orbit is less efficiently circularised by the ESF, as we investigate in more detail below.

Another good reason for considering ESF effects is to improve the modelling of astrophysical EMRIs. Firstly, compact bodies could sustain electromagnetic charges that could conceivably cause some dephasing over the final circa $10^4$--$10^5$ orbital cycles in the final year before plunge. Secondly, the presence of a dipole mode in the electromagnetic sector makes it a reasonable proxy for modelling EMRI evolution in beyond-GR theories. Theories with a radiative dipole that have been considered in the EMRI context include scalar-(vector)-tensor theories \cite{Khalil:2018aaj}, massive-gravity theories \cite{Cardoso:2023dwz}, and Einstein-Maxwell-Dilaton theory \cite{Hirschmann:2017psw}. Recent work by Barsanti \emph{et al.}~\cite{Barsanti:2022ana,Barsanti:2022vvl} has shown that, in in scenarios with a new fundamental scalar field, the secondary experiences a \emph{scalar} self-force of exactly the kind earlier considered as a toy model in the literature (e.g.~in Refs.~\cite{Warburton:2010eq,Warburton:2011hp}). It seems plausible (but is not shown here) that electromagnetic self force may play the same role in extended theories with a radiative spin-one degree of freedom. 

In the 1960s, the problem of a particle undergoing a gravitationally-driven inspiral into a black hole was first tackled using a Keplerian approximation to the underlying dynamics \cite{Peters:1963ux, Peters:1964zz}. The Keplerian approximation -- which can be thought of as the leading term in a post-Newtonian expansion -- gives a robust model for understanding the inspiral over much of its life. In this paper, we extend the Keplerian treatment of a binary black hole system to one where the particle's inspiral is driven by electromagnetism through the Abraham-Lorenz force, and then we proceed to a full-blooded radiation-reaction approach to calculate more accurate inspiral trajectories. 

The paper is organised as follows. In section \ref{sec:kepler}, we derive a Keplerian approximation to the problem of an electromagnetically-driven inspiral, and we recap the results in \cite{Peters:1963ux,Peters:1964zz} for a gravitationally-driven inspiral. In Sec.~\ref{sec:fluxes}, we summarise the calculation of the fluxes of energy and angular momentum radiated by particles on eccentric equatorial orbits in Kerr spacetime. The adiabatic inspiral model, in which the orbital parameters $(p,e)$ evolve in proportion to the fluxes, is described in Sec.~\ref{sec:InspiralModel}. In Sec.~\ref{sec:chebyshev}, we outline the application of Chebyshev interpolation to calculate fluxes across the $(p,e)$ parameter space from values at node points. Various implementation details are covered in Sec.~\ref{sec:implementation}. A selection of numerical results are presented in Sec.~\ref{sec:results}, including trajectories of inspirals in the $(p,e)$ plane. The article concludes in Sec.~\ref{sec:conclusions} with a recap of the key results, and an outlook on avenues for further exploration.

\section{Analysis\label{sec:analysis}}

 \subsection{Keplerian approximation\label{sec:kepler}}
For a Keplerian orbit, the position vector is $\mathbf{x}(t) = r(t) \cos \phi(t) \, \mathbf{i} + r(t) \sin \phi(t) \, \mathbf{j}$, where
\begin{align}
r(t) &= \frac{p}{1 + e \cos \phi(t)} , &
E &= -\frac{GMm}{2 \asm} , \\
 \dot{\phi} &= \frac{L}{r(t)^2},  & 
L &= \sqrt{GM p}. 
\end{align}
Here $p= \asm (1-e^2)$ is the semi-latus rectum, $\asm$ is the semi-major axis, and $e$ is the eccentricity of the orbit.
In the absence of other forces, the specific orbital energy $E$ and angular momentum $L$ are conserved, and the orbital parameters $(p,e)$ are constant. In the adiabatic approximation, the radiative losses cause $E$ and $L$ to change slowly (relative to the orbital timescale), and $\langle \dot{E} \rangle$ and $\langle \dot{L} \rangle$ are found by taking an average over one orbit.

\subsubsection{Electromagnetic-driven inspirals}

Invoking the Abraham-Lorentz \cite{abraham1905theorie,Lorentz:1892} force law, 
\beq
\mathbf{F} = \frac{2}{3} \frac{q^2}{4 \pi \epsilon_0 c^3} \dot{\mathbf{a}} , \label{eq:abraham-lorentz}
\eeq
where $\mathbf{v} = \dot{\mathbf{x}}$ and $\mathbf{a} = \dot{\mathbf{v}}$, the (instantaneous) rate of loss of energy and angular momentum are given by $\dot{E} = -\mathbf{v} \cdot \mathbf{F}$ and $\dot{L} = -\mathbf{x} \times \mathbf{F}$, respectively. Taking an average over one orbit, 
\begin{subequations}
\begin{align}
\langle \dot{E} \rangle &= - q^2 \frac{2}{3 \asm^4} \frac{(1 + \tfrac{1}{2} e^2)}{(1 - e^2)^{5/2}} , \label{eq:adot-kepler} \\
\langle \dot{L} \rangle &= - q^2 \frac{2}{3 \asm^{5/2} (1-e^2)} .
\end{align}
\end{subequations}
Here we have adopted units in which $GM = 1$, $m=1$, $c = 1$ and $4 \pi \epsilon_0 = 1$. 
Consequently,
\begin{subequations}
\begin{align}
\langle \dot{a} \rangle &= - \frac{4 q^2}{3 \asm^2} \frac{(1 + \tfrac{1}{2} e^2)}{(1 - e^2)^{5/2}} ,  \\ 
\langle \dot{e} \rangle &= - e \frac{q^2}{\asm^3 (1 - e^2)^{3/2}} . 
\end{align}
\end{subequations}
and so
\beq
\left\langle \frac{d \asm}{d e} \right\rangle  = \frac{4\asm}{3e} \frac{(1 + \tfrac{1}{2} e^2)}{1 - e^2} . 
\eeq
This ODE has an elementary solution,
\beq
p(e) = p_i \, \left( \frac{e}{e_i} \right)^{4/3} ,  \label{eq:pe-em}
\eeq
where $p_i$ and $e_i$ are initial values. Hence for a Schwarzschild black hole, the eccentricity at plunge, $e_f$, may be estimated from the iterative formula
\beq
 e_f = e_i \left( \frac{(6 + 2e_f) M}{p_i} \right)^{3/4} .  \label{eq:ef-at-plunge}
\eeq
The characteristic timescale for a circular-orbit inspiral is, from Eq.~(\ref{eq:adot-kepler}),
$
T = p_i^3 / (4 q^2) . 
$

\subsubsection{Gravitational-wave-driven inspirals}
For comparison with the above, we here quote the corresponding formulae for inspirals driven by gravitational radiation, also in the Keplerian approximation (see Refs.~\cite{Peters:1963ux,Peters:1964zz}), and for $m \ll M$. The rate of change of orbital parameters is
\begin{align}
\langle \dot{\asm} \rangle &= - \frac{64 \gamma}{5 \asm^3 (1 - e^2)^{7/2}} \left(1 + \frac{73}{24}e^2 + \frac{37}{96} e^4 \right)  ,  \\ 
\langle \dot{e} \rangle &= - \frac{304 e \gamma}{15 \asm^4 (1 - e^2)^{5/2}} \left(1 + \frac{121}{304} e^2 \right) ,  
\end{align}
where $\gamma = G^3 m_1 m_2 (m_1 + m_2) / c^5$, 
and so
\beq
\left\langle \frac{d \asm}{d e} \right\rangle  = \frac{12 \asm}{19 e} \frac{[1 + (73/24)e^2 + (37/96)e^4]}{(1-e^2) [1 + (121/304)e^2]}
\eeq
with the solution
\beq
p(e) = c_0 e^{12/19} \left( 1 + \frac{121}{304} e^2 \right)^{870/2299} , \label{eq:pe-grav}
\eeq
where $c_0$ is a constant of integration. 

By comparing the index $4/3$ in Eq.~(\ref{eq:pe-em}) with the index $12/19$ in Eq.~(\ref{eq:pe-grav}), it is clear that the emission of gravitational waves will ``circularize'' the orbit more efficiently than the emission of electromagnetic waves (at least within this approximation). 

 \subsection{Flux calculation on Kerr spacetime\label{sec:fluxes}}
The approximations derived in the previous direction are valid in the far-field, $p \gg M$. For a more accurate study, it is necessary to compute the fluxes for orbits in the strong-field region, all the way up to the plunge. This is done by applying the Teukolsky formalism \cite{Teukolsky:1972my,Teukolsky:1973ha,Press:1973zz,Teukolsky:1974yv, chandrasekhar1976solution,chandrasekhar1998mathematical}. 

For an equatorial bound geodesic on Kerr spacetime, the averaged fluxes of energy ($E$) and angular momentum ($L$) at infinity, and at the horizon, are (see Appendix \ref{appendix:fluxes} for a derivation),
\begin{subequations}
\begin{align}
\Phi^{(E)}_{\infty} &= \frac{1}{8 \pi} \sum_{l m n} \left| \alpha_{-1}^{\infty} \right|^2 , &  \label{eqn:EnergyFluxes}
\Phi^{(L)}_{\infty} &= \frac{1}{8 \pi} \sum_{l m n} \frac{m}{\omega} \left| \alpha_{-1}^{\infty} \right|^2 , \\
\Phi^{(E)}_{h} &= \frac{1}{8 \pi} \sum_{l m n} \frac{\omega}{2 M r_+ \tilde{\omega}} \left| \alpha_{+1}^{h} \right|^2 , &
\Phi^{(L)}_{h} &= \frac{1}{8 \pi} \sum_{l m n} \frac{m}{2 M r_+ \tilde{\omega}} \left| \alpha_{+1}^{h} \right|^2 , \label{eqn:MomentumFluxes}
\end{align}
\label{eq:fluxes}
\end{subequations}
where  $\omega = \omega_{mn} \equiv m \Omega_\phi + n \Omega_r$ and $\tilde{\omega} = \omega - m \Omega_H$, with $\Omega_H = a / (2Mr_+)$ the angular frequency of the horizon. The integers $\ell m n$ take values in the ranges $\ell \in \{0, 1, \ldots \infty\}$, $m \in \{-\ell, -\ell+1, \ldots, \ell\}$ and $n \in \{-\infty, \ldots,-1,  0 , 1 , \ldots \infty \}$. The coefficients are
\begin{align}
\alpha^{\infty}_{\pm 1} &= \frac{4 \pi q}{\sqrt{2}}  \frac{1}{T_r} \int_0^{T_r} \frac{dt}{u^t} \frac{e^{i \omega_{mn} t_0 - i m \phi_0}}{r_0 W_{\pm}} \left\{ \left( (A_{\pm}(r_0) S_{\pm1}(\tfrac{\pi}{2}) - u_{l^\pm}^{(0)}  S'_{\pm1}(\tfrac{\pi}{2}) \right) P_{\pm 1}^h(r_0) + u_{m^\pm}^{(0)} S_{\pm1}(\tfrac{\pi}{2}) P_{\pm 1}^{h \prime}(r_0) \right\} ,   \label{eq:alpha-inf}
\end{align}
with 
\begin{align}
A_{\pm}(r_0) &= u_{l^\pm}^{(0)} \left( \mp Q_0 + \frac{i a}{r_0} \right) - u^{(0)}_{m^{\pm}} \left(\mp \frac{i K_0}{\Delta_0} + \frac{1}{r_0} \right) , \label{eq:Adef}
\end{align}
where $Q_0 = m - a \omega$ and $K_0 = \omega (r_0^2+a^2) - am$, and
\begin{subequations}
\begin{align}
u_{l^\pm}^{(0)} &= l_{\pm}^\mu u_\mu = \frac{r_0^2 \dot{r}_0 \mp (r_0^2 + a^2) E \pm a L}{\Delta_0} , \\
u_{m^\pm}^{(0)} &= m_{\pm}^\mu u_\mu = \pm i (L - a E) .
\end{align} 
\label{u0def}
\end{subequations}
The functions $P_{\pm1}^{h/\infty}(r)$ are (rescaled) Teukolsky radial functions for spins $\pm1$ satisfying IN ($h$) and UP ($\infty$) boundary conditions, and $W_{\pm} \equiv P_{\pm1}^{h} \partial_r P_{\pm1}^{\infty} - P_{\pm1}^{\infty} \partial_r P_{\pm1}^{h} $ is their Wronskian; the functions $S_{\pm1}(\theta)$ are spheroidal harmonics of spin-weight $\pm 1$. 
The coefficients $\alpha_{\pm1}^{h}$ are found by switching $\infty$ and $h$ in Eq.~(\ref{eq:alpha-inf}) (i.e.~switching the UP and IN homogeneous modes). The steps in the derivation of Eq.~(\ref{eq:alpha-inf}) are contained in Appendix \ref{appendix:fluxes}.

\subsection{Adiabatic inspiral model \label{sec:InspiralModel}}

 In the adiabatic approximation, we assume that the particle's worldline is, at each moment, locally tangent to an equatorial eccentric geodesic with orbital parameters $p$ and $e$ (the semi-latus rectum and eccentricity, respectively). These orbital parameters change smoothly with time -- and slowly in comparison with the orbital timescale --- as the system radiates away energy and angular momentum \cite{Osborn2015}. Thus each inspiral corresponds to a trajectory in the $(p,e)$-plane. For a particle on an eccentric orbit, the radius of orbit $r_0$ is related to the orbit's semi-latus rectum $p$ and eccentricity $e$ by
\begin{equation}
    r_0(\tau) = \frac{pM}{1+e\cos(\chi(\tau))}, \label{eqn:r(e,p)}
\end{equation}
where $\chi$ is a monotonically increasing function of proper time $\tau$. 

The energy and angular momentum of a particle on an equatorial (geodesic) orbit can be expressed as functions of the orbital parameters, as $E=E(p,e)$ and $L=L(p,e)$. The form of these functions is specified in Appendix \ref{appendix:ELpe}. Thus, we can relate $\dot{E}$ and $\dot{L}$ to $\dot{p}$ and $\dot{e}$ by
\begin{align}
    \dot{E} &= \pdv{E}{p} \dot{p} + \pdv{E}{e} \dot{e}, \\
    \dot{L} &= \pdv{L}{p} \dot{p} + \pdv{L}{e} \dot{e}, 
\end{align}
i.e., a system of differential equations that can be written in matrix form and inverted giving
\begin{equation}
    \mqty(\dot{p}\\\dot{e}) = \frac{1}{\pdv{E}{p} \pdv{L}{e} - \pdv{L}{p}\pdv{E}{e}} \mqty(\pdv{L}{e} & - \pdv{E}{e} \\ - \pdv{L}{p} & \pdv{E}{p}) \mqty(\dot{E} \\ \dot{L}) , \label{eq:odesys}
\end{equation}
where the partial derivatives are calculated from Eq.~\eqref{Eq:Eep} and Eq.~\eqref{Eq:Lep}. Thus $\dot{E}$ and $\dot{L}$ determine how $e$ and $p$ change. If we specify $\dot{E}$ and $\dot{L}$, along with a set of initial conditions, we can solve this system of equations with an ODE integrator, such as Mathematica's \verb|NDSolve[]|.

Using a flux-balancing argument, in the adiabatic approximation the loss of orbital energy (orbital angular momentum) matches the averaged radiated energy flux (angular momentum flux). Hence the quantities $\dot{E}$ and $\dot{L}$ are given by the corresponding fluxes produced:
\begin{subequations}
\begin{align}
   - \dot{E} = \Phi^\infty_E(p,e) +\Phi^h_E(p,e) , \label{eqn:EdotFlux}\\
   - \dot{L} = \Phi^\infty_L(p,e) +\Phi^h_L(p,e) . \label{eqn:LdotFlux}
\end{align}
\label{eqn:ELdotFlux}
\end{subequations}
This completes the scheme for being able to calculate inspiral trajectories in the $(p,e)$ domain.

\subsection{Chebyshev interpolation\label{sec:chebyshev}}
To model an inspiral, we need fluxes across a domain in $(p,e)$ for a given spin parameter $a$. The numerical evaluation of $\Phi_{\infty/h}^{(E/L)}$ in Sec.~\ref{sec:fluxes} comes with a computational cost. We wish to be able to calculate these at any arbitrary point on the $(p,e)$ plane rapidly, so as to be able to calculate multiple trajectories. An efficient way to achieve this is by using interpolation with Chebyshev nodes to generate a polynomial that approximates the fluxes, as in Ref.~\cite{Lynch:2021ogr}. The resulting interpolation polynomial minimises the effect of Runge's phenomenon (i.e.~it minimises spurious oscillations near the boundaries of the interpolated region).

A Chebyshev polynomial of the first kind is a polynomial $T_n(x)$ such that $T_n(\cos \theta) = \cos(n\theta)$. The roots of these polynomials $x_k$ are known as Chebysehev nodes and are given by $x_k = \cos (\frac{2k-1}{2n} \pi)$ where $n$ is the degree of the polynomial and $k$ is an integer between 1 and $n$.
For our application, we wish to formulate a Chebyshev interpolation function for a function of two variables, $f(x,y)$. We wish to obtain an approximation of the form:
\begin{equation}
    f(x,y) \approx \sum\limits_{i,j} f^{ij} T_i(u(x)) T_j(v(y))\label{Eq:2DChebyshev}
\end{equation}
where $u$ and $v$ are scaled versions of our variables $x,y$ such that $u,v\in [-1,1]$ and $f^{ij}$ are the associated Chebyshev coefficients. The procedure for calculating these coefficients is:
\begin{enumerate}
    \item Set up an $n\times m$ grid of Chebyshev nodes $(x_i, y_j)$ on the domain of interest. 
    \item At each node calculate the value of the function $f(x_i,y_j)$, and plug into Eq.~\eqref{Eq:2DChebyshev} giving:
    \begin{equation}
        f(x_i,y_j) = \sum\limits_{i=0}^{n-1}\sum\limits_{j=0}^{m-1} f^{ij} T_i(u(x_i))T_j(v(y_j)) 
    \end{equation}
    This is a set of $n\times m$ equations for $n \times m$ unknowns. 
    \item Solve this system of equations using Mathematica's \verb|NSolve[]| function to obtain $f^{ij}$.
\end{enumerate}

\subsection{Implementation details\label{sec:implementation}}
\begin{figure}[h]
   \begin{center}
    \begin{subfigure}[b]{0.45\textwidth}
        \includegraphics[width = \textwidth]{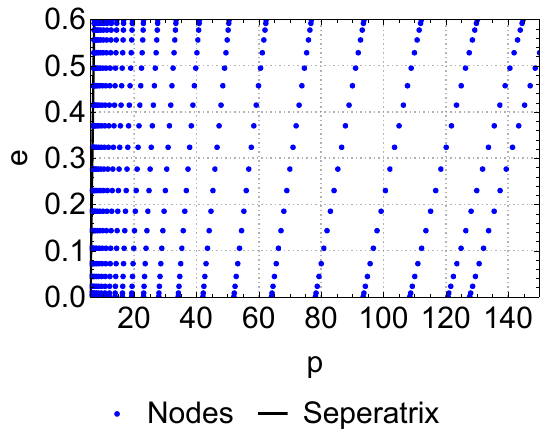}
        \caption{Full range of nodes.}
        %\label{}
    \end{subfigure}
    \hfill
    \begin{subfigure}[b]{0.45\textwidth}
        \includegraphics[width = \textwidth]{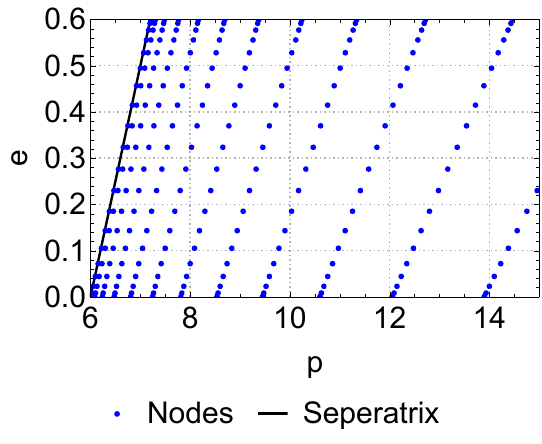}
        \caption{Nodes Close to the Separatrix.}
        %\label{}
    \end{subfigure}
   \end{center}
    \caption{Illustrating the $25 \times 20$ grid of nodes chosen for interpolating the fluxes in the $(p,e)$ plane ($a=0$ case).}
    \label{fig:nodes}
\end{figure}

These methods were implemented in Mathematica. A $25\times20$ grid of nodes was chosen to perform the interpolation. We define a parameter $y = \sqrt{p_\text{separ}(a,e) / p}$. This transforms the $pe$-plane into the $ye$-plane.  We do this transformation to only calculate nodes beyond the Separatrix, the line in the $pe$-plane that separates orbits from plunging trajectories. We then map this plane onto the $[-1,1]^2$ domain upon which the Chebyshev polynomials are defined using the following transformations:
\begin{align}
    u = \frac{y - (y_\text{min} + y_\text{max})/2}{(y_\text{min} - y_\text{max})/2} && v = \frac{e - (e_\text{min} + e_\text{max})/2}{(e_\text{min} - e_\text{max})/2}
\end{align}
where the subscripts indicate the minimum and maximum values we want to consider on the plane. We choose $e_\text{min} = 0$ and $e_\text{max} = 0.6$, and $y_\text{max}=1$,  corresponding to the point on the separatrix, and  $y_\text{min}= \sqrt{p_\text{separ}(a,0.5) / 150}$, which is a value of $y$ that roughly corresponds with $p=150$ as a maximum $p$ value. We then define the Chebyshev nodes on this new $uv$-plane and invert the transformations to obtain the nodes in the $pe$-plane, these are illustrated in Fig.~\ref{fig:nodes}. 

\begin{figure}
    \centering
    \includegraphics[width = 0.8\textwidth]{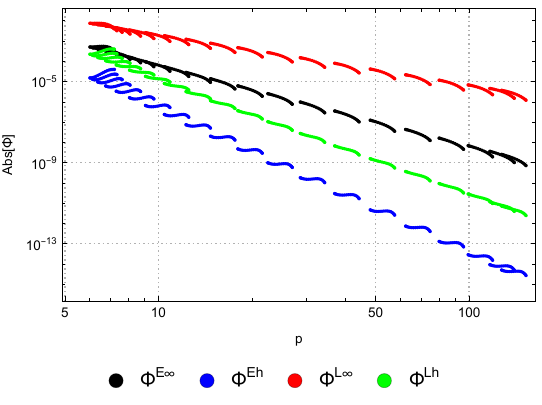}
    \caption{The absolute magnitude of fluxes as a function of semi-latus rectum $p$, for eccentricities $e \in [0, 0.6]$. Each line is increasing in $e$ from left to right. Shown are the energy and angular momentum flux at infinity (black and red) and at the horizon (blue and green). 
    }
    \label{fig:loglogplt}
\end{figure}

In principle, the total fluxes in Eq.~\eqref{eqn:EnergyFluxes} and Eq.~\eqref{eqn:MomentumFluxes} are formed from infinite sums over $\ell m n$ modes. These partial fluxes are typically monotonically decreasing in magnitude with $|n|$, and diminish in an exponential fashion. Thus when calculating the final flux we can truncate the sum to some $N_\text{max}$ and $L_\text{max}$ values without significant loss of accuracy. 
Furthermore, there is a symmetry that can be exploited to increase calculation speed two-fold. For a given flux mode $F_{\ell mn}$ we have that $F_{\ell m n} = F_{\ell -m -n}$, and that $F_{\ell 0 0} = 0$. Applying this symmetry to the sum over flux modes we get that the total flux $\Phi$ will be given by:
\begin{equation}
    \Phi = 2 \sum\limits_{\ell= 1} ^{L_\text{max}} \left[\sum\limits_{n=0}^{N_\text{max}} \left(F _{\ell 0 n} + \sum\limits_{m=1}^{\ell} (F_{\ell m n} + F_{\ell m -n})\right) - \sum_{m=1}^{\ell} F_{\ell m 0}\right] . \label{eqn:SymmetryEvaluationFlux}
\end{equation}
Each of the four fluxes were calculated at each node point for $a/M = 0, 0.25, 0.5, 0.75, 0.9$, and $q=1$. Once calculated, we plotted each set of flux points against its $p$-value. This is illustrated in Fig.~\ref{fig:loglogplt}. We can fit a straight line to each of these, the coefficient of which can be used as a smoothing parameter on the flux to give a flatter function before applying the Chebyshev interpolation. Numerically, we obtain smoothing parameters $(s_{E\infty}, s_{Eh}, s_{L\infty}, s_{Lh}) = (4.16, 6.53, 2.68, 5.14)$ for $a = 0$, which are close to the indices anticipated from the post-Newtonian expansions. We then define the function we will interpolate as $f(p,e) = p^{s_{(i)}} \Phi^{(i)}(p,e)$, and then the final interpolated fluxes $\Phi^{(i)}_\text{interp} = f(p,e) / p^{s_{(i)}}$, where $(i)$ indicates the type of flux.

With all this in place we can now calculate inspiral trajectories by feeding Eqs.~(\ref{eqn:ELdotFlux}) with the interpolated fluxes, and solving the ODE system (\ref{eq:odesys}) numerically using \verb|NDSolve[]|.

\section{Results\label{sec:results}}

\begin{figure}
    \centering
    \includegraphics[width = 1\textwidth]{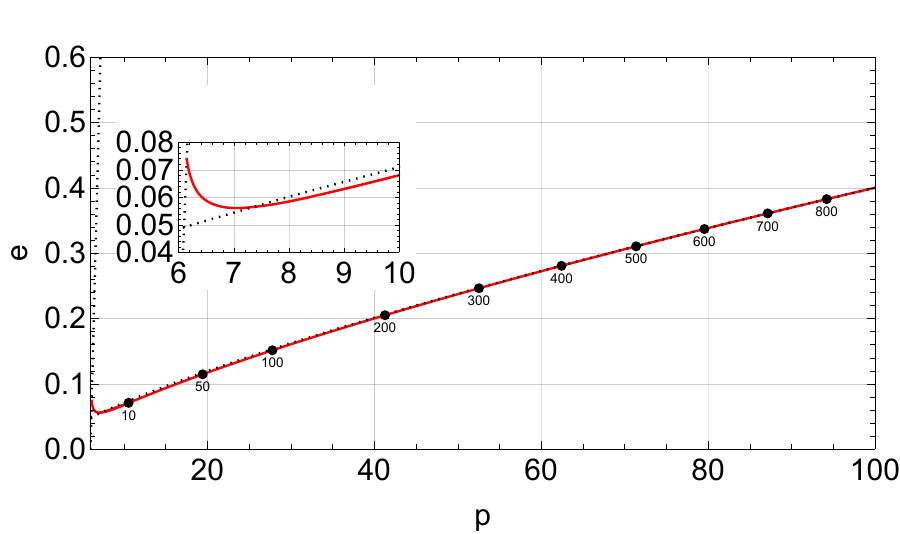}
    \caption{Inspiral trajectory in the $(p,e)$ plane for $p_i=100$, $e_i =0.4$ and $a = 0$. The vertical dotted line is the separatrix, marking the onset of the plunge phase. The horizontal dotted line is the Keplerian approximation, Eq.~(\ref{eq:pe-em}). The labels on the plot indicate the number of orbits remaining until plunge, in the case $q/m = 1$ (see Appendix \ref{appendix:orbits}). 
    }
    \label{fig:InspiralTrajectory}
\end{figure}

Figure \ref{fig:InspiralTrajectory} shows the calculated inspiral trajectory in the $(p,e)$ plane for initial values $e_i=0.4$, $p_i=100$ in the Schwarzschild case ($a=0$). In the far field ($p \gg M$) there is excellent agreement between the Keplerian approximation and the numerical results obtained from the flux-balancing argument. As the particle gets close to the black hole, some deviation begins to occur, with the trajectory first dipping below the eccentricity predicted by the Keplerian approximation, before an uptick in eccentricity in the final stages of the inspiral before the plunge. The uptick in eccentricity is familiar from the gravitational case \cite{Lynch:2021ogr}. 

As expected, the inspiral speeds up as the particle gets closer to the black hole and the radiated fluxes increase in magnitude. This progression is visible in Fig.~\ref{fig:InspiralTrajectory} by the number of orbits before plunge (for the case $q/m=1$); the calculation of this quantity is described in Appendix \ref{appendix:orbits}.

Figure \ref{fig:a0vearyinge} shows electromagnetic inspiral trajectories for various initial eccentricities. In every case, the orbit becomes more circular as the inspiral progresses. However, circularization due to EM emission is less efficient than circularization due to gravitational-wave emission, and significant eccentricity is retained at the point of plunge. 

\begin{figure}
    \centering
    \includegraphics{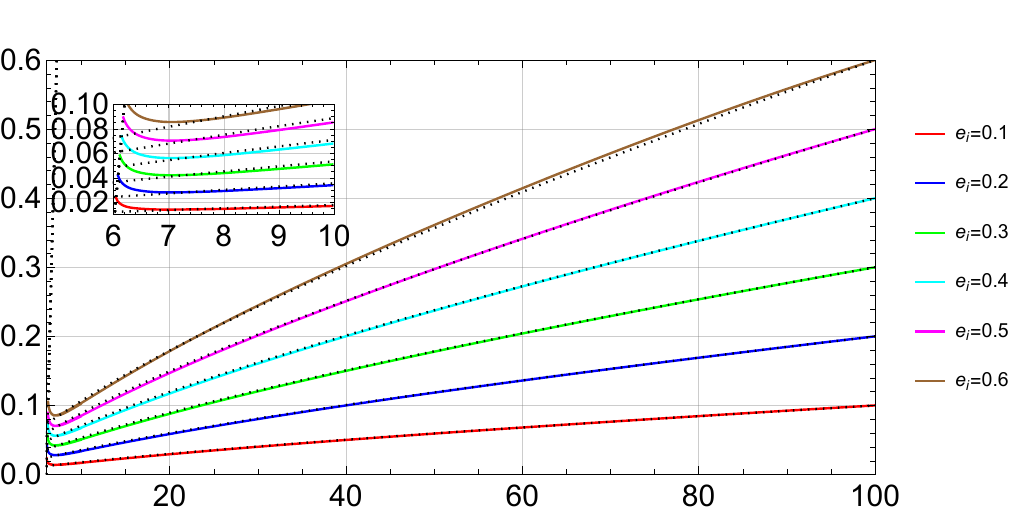}
    \caption{Inspiral trajectories in the $(p,e)$ plane for a selection of initial values of $e_i$ starting at $p_i = 100$ (Schwarzschild case $a=0$), comparing the adiabatic approximation (solid lines) with the Keplerian approximation (dotted lines). The inset shows the uptick in eccentricity as the trajectory approaches the separatrix. 
    }
    \label{fig:a0vearyinge}
\end{figure}

\begin{figure}
    \centering
    \includegraphics{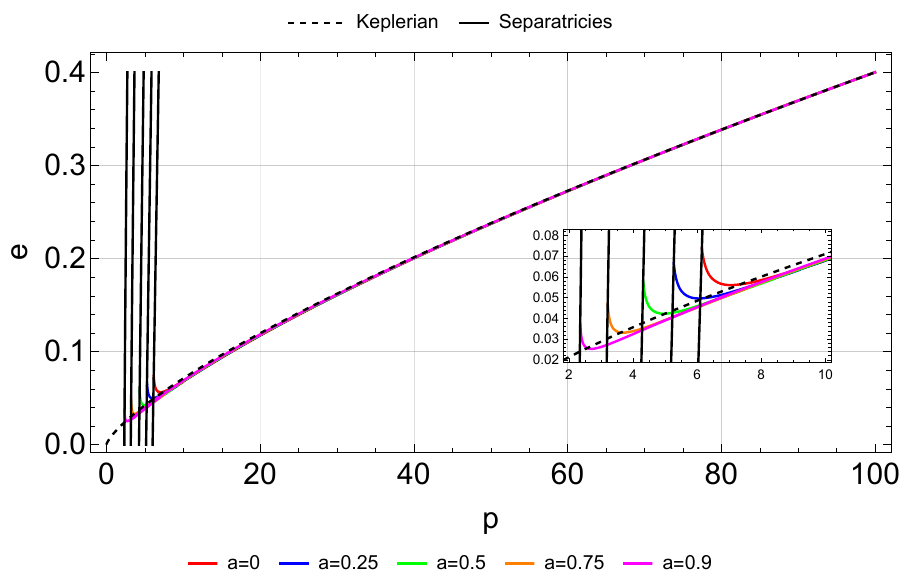}
    \caption{Inspiral trajectories in the $(p,e)$ plane for initial values $p_i = 100M$ , $e_i = 0.5$, for Kerr black holes of varying spin parameter $a/M$. The Keplerian approximation (\ref{eq:pe-em}) is shown as a dashed line. The inset shows the uptick in eccentricity in the approach to the separatrices (black solid lines). }
    \label{fig:VaryingA}
\end{figure}

Figure \ref{fig:VaryingA} shows inspiral trajectories for rotating (Kerr) black holes. The trajectories for $a \neq 0$ are qualitatively similar in nature to the Schwarzschild trajectory, again exhibiting a slightly faster loss of eccentricity than predicted by the Keplerian approximation \ref{eq:pe-em}, before an uptick in eccentricity in the approach to the plunge phase (i.e.~near the separatrix). The position of the separatrix depends on the spin of the black hole $a/M$. In the rapidly-spinning case, the co-rotating orbits make a closer approach to the black hole before plunging than the counter-rotating orbits.

\begin{figure}
    \centering
    \includegraphics{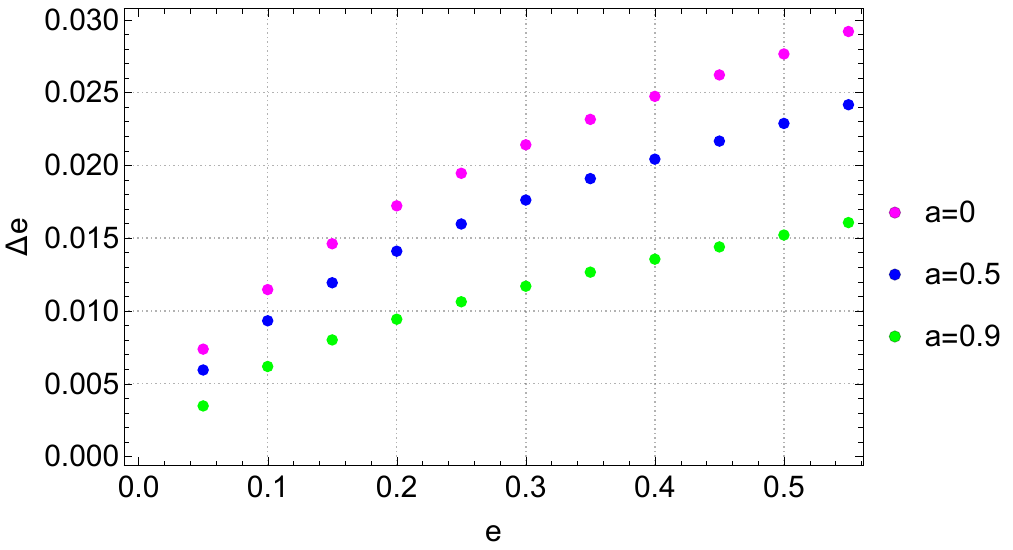}
    \caption{
    The difference $\Delta e \equiv e_{f} - e_{Kf}$ evaluated on the separatrix, where $e_f$ is evaluated on the adiabatic inspiral trajectory, and $e_{Kf}$ is evaluated with the Keplerian inspiral approximation (\ref{eq:ef-at-plunge}). This is shown as a function of the initial value of eccentricity $e_i$ at $p_i=100$, for black holes of spin $a/M \in \{0, 0.5, 0.9\}$ . 
    }
    \label{fig:diff-eccentricity}
\end{figure}

As shown above, the Keplerian approximation (\ref{eq:pe-em}) describes the inspiral accurately at large $p$, and robustly up to $p \sim 10$, but it fails to capture the uptick in eccentricity before the plunge. Consequently, Eq.~(\ref{eq:ef-at-plunge}) does not give a particularly accurate prediction of the eccentricity parameter at the plunge. Figure \ref{fig:diff-eccentricity} shows $\Delta e$ (defined in the caption), which characterises the additional eccentricity acquired in the uptick. As shown, the effect of the uptick diminishes somewhat as the spin of the black hole increases. 

\begin{figure}
    \centering
    \includegraphics{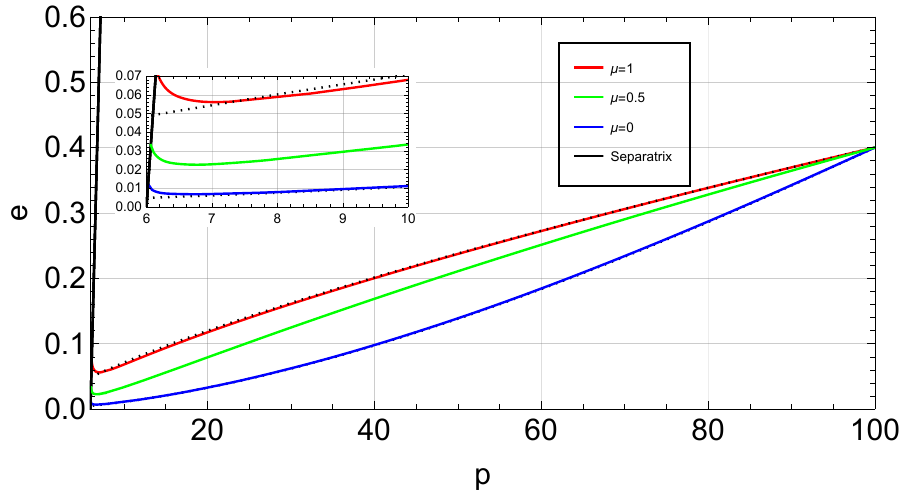}
    \caption{
    Inspirals driven by electromagnetic and gravitational fluxes, for initial conditions $(p_i,e_i)=(80 , 0.3)$ for a black hole with $M=1, a=0$. The dotted lines show the Keplerian approximations (\ref{eq:pe-em}) and (\ref{eq:pe-grav}). The mixing parameter $\mu$ is defined in Eq.~(\ref{eq:mu}).
    }
    \label{fig:EMGWInspirals}
\end{figure}

Inspirals in astrophysics are driven principally by the emission of gravitational waves, rather than electromagnetic waves. This motivates us to consider inspirals that are driven by both electromagnetic and gravitational interactions. We define the total \emph{mixed} flux as: 
\begin{equation}
    \Phi = \mu \Phi_\text{EM} + (1-\mu) \Phi_\text{G} , \label{eq:mu}
\end{equation}
where $\mu$ is a mixing parameter taking values on $[0,1]$, with $\mu=0$ corresponding to gravitational flux only and $\mu =1$ corresponding to electromagnetic flux only. Employing the mixed flux to calculate inspirals driven by the combined action of electromagnetism and gravity leads to the results shown in Fig.~\ref{fig:EMGWInspirals}. Here we see that the gravitationally-driven inspiral loses eccentricity faster than the electromagnetically-driven inspiral, as expected. All the trajectories show an uptick in the parameter $e$ in the approach to the separatrix.

\section{Discussion and conclusions\label{sec:conclusions}}

In the preceding sections, we have examined the adiabatic evolution of the orbital parameters of binary inspiral driven by  radiation reaction in electromagnetism. 
Figures \ref{fig:InspiralTrajectory}, \ref{fig:a0vearyinge}, \ref{fig:VaryingA}, \ref{fig:diff-eccentricity},  \ref{fig:EMGWInspirals} show trajectories in the $(p,e)$ plane, computed by interpolating the fluxes calculated through Teukolsky formalism (see Sec.~\ref{sec:fluxes}). 
We find that the leading-order Keplerian approximation, Eq.~(\ref{eq:pe-em}), gives a robust characterisation in the weak-field regime (Fig.~\ref{fig:InspiralTrajectory} and \ref{fig:a0vearyinge}). 
Near the plunge, there is a characteristic uptick in the eccentricity parameter $e$  (see Figs.~\ref{fig:a0vearyinge} and \ref{fig:diff-eccentricity}; a similar uptick is also observed in the gravitationally-driven case \cite{Cutler:1994pb}). 
The spin of the central black hole has some effect on the evolution of the inspiral in the strong-field regime, most notably in Fig.~\ref{fig:VaryingA} by shifting the innermost stable circular orbit that marks the transition to plunge (shifting inwards for prograde orbits, and outwards for retrograde orbits). 
As expected, the electromagnetic radiation-reaction causes the orbit to become less eccentric over time, but this circularization process is observed to be less efficient than in the gravitationally-driven case (Fig.~\ref{fig:EMGWInspirals}), which is consistent with the Keplerian approximation (see Eq.~(\ref{eq:pe-em}) and Eq.~(\ref{eq:pe-grav})). We expect this observation to hold true for any field theory with a radiative dipole mode.

% DeWitt, 1964: The questions answered by this investigation are of conceptual interest only, since the forces involved are far too small to be detected experimentally.

A few natural extensions of this work present themselves.
Firstly, it would be straightforward to include the interpolated EM flux model in the modular Fast EMRI Waveforms (FEW) framework \cite{Chua:2020stf}.
This would allow for fully Bayesian Markov Chain Monte Carlo-based parameter estimation studies in the time- \cite{Katz:2021yft} or frequency-domain \cite{Speri:2023jte} that could determine the smallest measurable charge on the secondary.
In fact, there has been recent work in this direction in Ref.~\cite{Zhang:2023vok} using \emph{circular} orbits, based on an assessment of the dephasing of gravitational waves over the lifetime of the binary. Extending such efforts to eccentric orbits is a natural next step.  

If residual charge were present in the binary, then it possible that the primary could also be charged, leading to a modified inspiral due to (i) direct EM interactions between the primary and secondary, and (ii) modifications of the spacetime due to the charge of the primary. Studying these effects in full would require the study of coupled EM and gravitational perturbations of the Kerr-Newman spacetime, which is a significant undertaking (see e.g.~\cite{Zhu:2018tzi}).

A second natural extension would be to consider non-equatorial orbits.
Very recently the \texttt{Teukolsky} package in the Black Hole Perturbation Toolkit was extended by one of us to compute the EM perturbation for a particle moving on generic bound orbit of a Kerr black hole.
Thus, computing the EM flux for spherical (fixed Boyer-Lindquist radius) or generic bound orbits is a straightforward, if computational demanding, task. 
Calculating the evolution of non-equatorial orbits adds a new complexity, as now the evolution of the Carter constant must also be obtained.
For spherical inspirals the zero-eccentricity constraint allows an evolution equation of the Carter constant to be obtained in terms of the asymptotic fluxes \cite{Hughes:1999bq}.
For generic inspirals (eccentric and inclined) no such balance law can be obtained in general. 
Fortunately, so long as the binary is evolving adiabatically, Mino showed how the correct radiation reaction force for scalar, EM, or gravitational perturbations can be derived \cite{Mino:2003yg}.
This result was then used by others to provide explicit, practical evolution equations for the Carter constant in the scalar \cite{Drasco:2005is} and gravitational \cite{Sago:2005fn,Sago:2005gd} cases.
To the best of our knowledge, work remains to be done to extend these results to the electromagnetic case.

Thirdly, there is clearly scope for obtaining post-Newtonian expansions of fluxes and/or the local self-force. From these, closed-form approximations for $p(e)$ would follow, to go substantially beyond the Keplerian approximation of Sec.~\ref{sec:kepler}. In the gravitational case, there has been impressive progress in calculating fluxes (e.g.~Refs.~\cite{Fujita:2014eta,Munna:2020iju,Munna:2023vds}), the self-force acting on the particle (e.g.~Ref.~\cite{Hopper:2015icj}), and certain gauge-invariant scalars (see e.g.~\cite{Bini:2018qvd, Munna:2022xts}) at very high orders in $v/c$, at leading order in $m/M$. The technical machinery for expanding solutions of the Teukolsky equations should transfer straightforwardly from the $s=2$ case to the $s=1$ case, and so the prospects for such calculations appear to be good. 

One limitation of the adiabatic approximation, employed here, is that \emph{conservative} effects of the self-force are neglected. At leading order in $r^{-1}$ the conservative part of the self-force is \cite{DeWitt:1964} $\mathbf{F}_{\text{cons}} \approx \frac{q^2}{4 \pi \epsilon_0 c^2} \cdot \frac{GM}{r^3} \hat{\mathbf{r}}$. The conservative component oscillates over one orbit, unlike the dissipative force which drives the secular change in $E$ and $L$. Including the conservative force is necessary for an accurate calculation of the orbital phase, and thus dephasing. An obvious next step is to calculate the local self-force on the charged particle using e.g.~the $\ell$-mode regularization scheme (as was done for Schwarzschild circular orbits in Ref.~\cite{Nolan-thesis}, Schwarzschild eccentric orbits in Ref.~\cite{Haas:2011np}, and Kerr circular orbits in Ref.~\cite{Torres:2020fye}).
This would enable the calculation of the shift in the innermost stable circular orbit (ISCO) under the conservative piece of the electromagnetic self-force; its gravitational counterpart is already well characterised \cite{Barack:2009ey, Isoyama:2014mja}. 

Finally, in this work we have restricted our attention to binaries where the primary spin $|a| \le 0.9M$.
Astrophysically, we might expect spins higher spins up to $|a| = 0.998M$. Beyond this, the near-horizon, near-extremal ($|a| > 0.999$) is also interesting as new physical effect occur in this regime. Previous work has both analytically and numerically calculated how the flux behaves in this case for an orbiting scalar and gravitational charge \cite{Gralla:2015rpa,Gralla:2016qfw}.
The calculations could be extended to the EM case.

\section*{Acknowledgements}

This work makes use of the \texttt{Black Hole Perturbation Toolkit} \cite{BHPToolkit}. In particular we make use of the \texttt{KerrGeodesics}~\cite{niels_warburton_2023_8108265},
\texttt{SpinWeightedSpheroidalHarmonics}~\cite{barry_wardell_2023_8112931}, and \texttt{Teukolsky}~\cite{barry_wardell_2023_8116897} packages. S.D.~thanks Martin van de Meent for his advice on the derivation in Appendix 
\ref{appendix:fluxes}. 

\appendix

\section{Fluxes from Teukolsky variables\label{appendix:fluxes}}
In this section we derive the expressions for the fluxes in Eqs.~(\ref{eq:fluxes}). For the gravitational case, a related derivation is found in Ref.~\cite{Hughes:1999bq}. As presented, the derivation below is not mathematically rigorous because we do not address issues of convergence of the various sums and integrals; we assume that any such issues can be handled by introducing regulators and limits. 

The starting point is the electromagnetic field equations, $\nabla_{\nu} F^{\mu \nu} = 4 \pi J^\mu$ and $\nabla_{[\mu} F_{\nu \sigma]} = 0$, which relate the Maxwell tensor $F^{\mu \nu}$ to the four-current $J^\mu$ sourced by a particle of charge $q$ on a worldline $x_0^\mu(\tau)$ with tangent $u^\mu \equiv dx_0^\mu / d\tau$, given by
\beq
J^\mu = q \int u^\mu \delta^4 (x - x_0(\tau)) d \tau .  \label{eq:Jmu}
\eeq
Here $\delta^4(\cdot)$ is the four-dimensional Dirac delta distribution, with support on the particle worldline.

Teukolsky \cite{Teukolsky:1972my,Teukolsky:1973ha} showed that on Kerr spacetime, the (scaled) Maxwell scalars of extremal spin-weight satisfy decoupled second-order equations which are separable. The equation for $\phi_0 \equiv F_{\mu \nu} l_+^\mu m_+^\nu$ takes the form \cite{Bini:2002jx}
\beq
 \left( \left( \nabla_\mu + \Gamma_\mu \right) \left( \nabla^\mu + \Gamma^\mu \right) - 4 \psi_2 \right) \phi_0 = 4 \pi  J_0 , \label{eq:4Dteuk}
\eeq
where $\psi_2 = M/(r - i a \cos \theta)^3$, and the ``connection vector'' $\Gamma^\mu$ is listed Eq.~(4.11) of Ref.~\cite{Bini:2002jx}. Here $J_0$ is a scalar source term derived from $J^\mu$ by taking certain first-order derivatives of the projections $J_l \equiv J_\mu l^\mu_+$ and $J_m \equiv J_\mu m^\mu_+$ (see below). This equation is separable, with the ansatz
\beq
\phi_0 = \int d \omega \sum_{\ell m} \Delta^{-1} P_{+1}^{\ell m \omega}(r) S_{+1}^{\ell m \gamma}(\theta) e^{i \chi} ,  \label{eq:phi0-expansion}
\eeq
where $\gamma = a \omega$ and $\chi = - \omega t + m \phi$, leading to a set of radial and angular equations,
\begin{align}
\left(\Delta \cD^\dagger \cD + 2 i \omega r - \lambda \right) P_{+1}(r) &= \mathcal{T}_{+1}(r) , \label{eq:radial-sourced} \\
\left(\cL^\dagger (\cL + \cot\theta) - 2 a \omega \cos \theta + \lambda \right) S_{+1}(\theta) &= 0 .
\end{align}
Here $P_{+1}(r) = P_{+1}^{\ell m \omega}(r)$ and $S_{+1}(\theta) = S_{+1}^{\ell m \gamma}(\theta)$ are radial and angular Teukolsky functions with $s=+1$; $\lambda = \lambda_{-1}^{\ell m \gamma}$ is the angular separation constant for $s=-1$; and $\mathcal{T}_{+1}(r)$ is a source term to be derived from $J_0$ (see below). Henceforth the mode labels $\ell m \omega \gamma$ are typically omitted for brevity. The operators $\hat{\cD} \equiv l_+^\mu \partial_\mu$, $\hat{\cD}^\dagger \equiv l_-^\mu \partial_\mu$, $\hat{\cL}^\dagger \equiv m_+^\mu \partial_\mu$ and $\hat{\cL} \equiv m_-^\mu \partial_\mu$ are Chandrasekhar's directional derivatives \cite{chandrasekhar1976solution,chandrasekhar1998mathematical}, and here these always act on terms with $e^{i \chi}$ dependence, so that $\hat{\cD} (e^{i \chi} \mathcal{X}(r,\theta)) = e^{i \chi} \cD \mathcal{X}(r,\theta)$, etc., with
\begin{subequations}
\begin{align}
\cD  &= \partial_r - \frac{i K}{\Delta_r},  & 
\cL^\dagger &=  \partial_\theta - Q ,
 \\
\cD^\dagger  &= \partial_r + \frac{i K}{\Delta_r} , 
& \cL &= \partial_\theta + Q ,
\end{align}
\label{eq:DLoperators}
\end{subequations}
and $K \equiv \omega (r^2 + a^2) - a m$ and $Q \equiv m \csc \theta - a \omega \sin \theta$. 

To proceed further, we require an expression for the radial source term $\mathcal{T}_{+1}(r)$ for a particle on an equatorial orbit. We start with the four-current (\ref{eq:Jmu}), and the delta-distribution taking the form
\beq
\delta^4(x - x_0) \equiv \frac{1}{\sqrt{-g}} \prod_\mu \delta(x^\mu - x^\mu_0) = \frac{1}{r_0^2} \delta(t - t_0) \delta(r-r_0) \delta(\theta - \pi/2) \delta(\phi - \phi_0). 
\eeq
We can decompose the source term into $\omega m$ modes by using standard representations of the delta functions,
\begin{align}
\delta (t - t_0) &= \frac{1}{2 \pi} \int d \omega e^{- i \omega (t - t_0)} , \\
\delta (\phi - \phi_0) &= \frac{1}{2 \pi} \sum_m e^{i m (\phi - \phi_0) } .
\end{align}
From the definition of $J_0$ in Ref.~\cite{Teukolsky:1973ha}, and the properties of the delta distribution, it follows that the source in Eq.~(\ref{eq:4Dteuk}) takes the form
\beq
\Sigma J_0 = \frac{1}{(2 \pi)^2} \frac{q}{\sqrt{2}} \int d \omega \sum_m e^{i \chi} \int \frac{d \tau}{r_0} e^{- i \chi_0} \left[ u_{l^+}^{(0)} \left( \cL^\dagger + \frac{ia}{r_0} \right) - u_{m^+}^{(0)} \left(\cD + \frac{1}{r_0} \right) \right] \delta(r - r_0) \delta(\theta - \pi/2) ,
\eeq
with $\chi_0 = - \omega t_0 + m \phi_0$, and $u_{l^+}^{(0)}$ and  $u_{m^+}^{(0)}$ as defined in Eq.~(\ref{u0def}). At this point, we can use the completeness of the basis of spheroidal harmonics \cite{Breuer1977} to make the replacements 
\begin{align}
\delta(\theta - \pi/2) &= 2 \pi \sum_\ell S_{+1}^{\ell m \gamma}(\theta) S_{+1}^{\ell m \gamma}(\pi/2) , \\
\frac{\partial}{\partial \theta} \delta(\theta - \pi/2) &= - 2 \pi \sum_\ell S_{+1}^{\ell m \gamma}(\theta) S_{+1}^{\ell m \gamma \prime}(\pi/2) .
\end{align}
Consequently we arrive at an expression for the radial source function in Eq.~(\ref{eq:radial-sourced}), 
\beq
\mathcal{T}_{+1} = \sqrt{2} q \Delta \int \frac{d \tau}{r_0}  e^{- i \chi_0}  \left\{ \left( A(r_0) S_{+1}(\tfrac{\pi}{2}) - u_{l^+}^{(0)} S_{+1}^\prime(\tfrac{\pi}{2}) \right)  \delta(r - r_0) - u_{m^+}^{(0)} S_{+1}(\tfrac{\pi}{2}) \delta^\prime(r - r_0)  \right\} ,
\eeq
where $A(r_0)$ is defined in Eq.~(\ref{eq:Adef}). 
The ODE in Eq.~(\ref{eq:radial-sourced}) can be solved by first constructing the Green's function constructed from the IN and UP homogeneous solutions, $P^h_{+1}(r)$ and $P^\infty_{+1}(r)$, as
\beq
g_{+1}(r,r') \equiv \frac{1}{\Delta(r') W}  \left( P^{\infty}_{+1}(r) P^{h}_{+1}(r') \Theta(r - r') + P^{h}_{+1}(r) P^{\infty}_{+1}(r') \Theta(r' - r) \right) ,
\eeq
where $W = W_{+}$ is the (constant) Wronskian defined below Eq.~(\ref{u0def}). 
The inhomogeneous solution is
\beq
P_{+1}(r) = \int g_{+1}(r,r') \mathcal{T}_{+1}(r') dr' = \mathcal{C}^{\infty}_{+1}(r) P^{\infty}_{+1}(r) + \mathcal{C}^{h}_{+1}(r) P^{h}_{+1}(r) 
\eeq
where
\begin{align}
\mathcal{C}^{\infty}_{+1}(r) &\equiv \int_{\rmin}^r \frac{P_{+1}^h(r') \mathcal{T}_{+1}(r')}{\Delta(r') W_{+1}} dr' , &
\mathcal{C}^{h}_{+1}(r) &\equiv \int_{r}^{\rmax} \frac{P_{+1}^{\infty}(r') \mathcal{T}_{+1}(r')}{\Delta(r') W_{+1}} dr' , 
\end{align}
and $\rmin$/$\rmax$ are periapsis/apoapsis radial coordinates of the bound orbit. Here $\mathcal{C}^{\infty}_{+1}(r) = 0$ for $r < \rmin$ and $\mathcal{C}^{h}_{+1}(r) = 0$ for $r > \rmax$. Consequently, outside the libration region,
\beq
P_{+1}(r) = \begin{cases} C_{+1}^{\infty} P_{+1}^{\infty}(r), & r \ge \rmax , \\
C_{+1}^{h} P_{+1}^{h}(r), & r \le \rmin , \end{cases} \label{eq:Pasymptotics}
\eeq
where $C^{\infty}_{+1} \equiv \mathcal{C}^{\infty}_{+1}(\rmax)$ and $C^{h}_{+1} \equiv \mathcal{C}^{h}_{+1}(\rmin)$. To evaluate these coefficients, we may swap the order of the $\tau$ and $r$ integrals, and integrate out the $\delta$-distributions, to obtain a smooth function of $t_0$ for the integrand. That is, 
\begin{align}
C^{\infty}_{+1} &= \frac{\sqrt{2} q}{W} \int_{\rmin}^{\rmax} dr P ^{h}_{+1}(r) \int \frac{d \tau}{r_0} e^{-i \chi_0} 
 \left\{ \left( A(r_0) S_{+1}(\tfrac{\pi}{2}) - u_{l^+}^{(0)} S_{+1}^\prime(\tfrac{\pi}{2}) \right)  \delta(r - r_0) - u_{m^+}^{(0)} S_{+1}(\tfrac{\pi}{2}) \delta^\prime(r - r_0)  \right\} , \\
 &= \frac{\sqrt{2} q}{W} \int \frac{d t_0}{u^t r_0} e^{-i \chi_0}  \left\{ \left( A(r_0) S_{+1}(\tfrac{\pi}{2}) - u_{l^+}^{(0)} S_{+1}^\prime(\tfrac{\pi}{2}) \right) P^h_{+1}(r_0) + u_{m^+}^{(0)} S_{+1}(\tfrac{\pi}{2}) P^{h\prime}_{+1}(r_0) \right\} .
\end{align}
This can be expressed in the following form,
\begin{align}
C^{\infty}_{+1} &= \int dt_0 e^{i (\omega - m \Omega_\phi) t_0} \mathfrak{C}(t_0) ,  \label{eq:int-frakC} \\
\mathfrak{C}(t_0) &=  \frac{\sqrt{2} q}{W u^t r_0} e^{-i m (\phi_0 - \Omega_\phi t_0)}  \left\{ \left( A(r_0) S_{+1}(\tfrac{\pi}{2}) - u_{l^+}^{(0)} S_{+1}^\prime(\tfrac{\pi}{2}) \right) P^h_{+1}(r_0) + u_{m^+}^{(0)} S_{+1}(\tfrac{\pi}{2}) P^{h\prime}_{+1}(r_0) \right\} ,
\end{align}
where $\mathfrak{C}(t_0)$ is periodic, such that $\mathfrak{C}(t_0 + n T_r) = \mathfrak{C}(t_0)$, where $T_r = 2\pi / \Omega_r$ is the radial period of the eccentric orbit. Exploiting this periodicity, we can substitute the Fourier-series representation, 
\begin{align}
\mathfrak{C}(t_0) &= \sum_{n} \widetilde{\mathfrak{C}}_n e^{- i n \Omega_r t_0} , &
\widetilde{\mathfrak{C}}_n \equiv \frac{1}{T_r} \int_0^{T_r} dt_0 \mathfrak{C}(t_0) e^{i n \Omega_r t_0},
\end{align}
into the integral (\ref{eq:int-frakC}), to obtain
\begin{align}
C^{\infty}_{+1} &= \sum_{n} 2 \pi \, \widetilde{\mathfrak{C}}_n \, \delta \left( \omega - \omega_{mn} \right) . \label{eq:Cspikes}
\end{align}
Inserting (\ref{eq:Cspikes}) into (\ref{eq:Pasymptotics}) and (\ref{eq:phi0-expansion}), and performing the integral over frequency, yields a mode-sum for the Maxwell scalar outside the libration region, 
\beq
\phi_0 = \Delta^{-1} \sum_{\ell m n} S_{+1}^{\ell m \gamma}(\theta) e^{- i \omega_{mn} t + i m \phi}  
\begin{cases}
 \alpha_{+1}^{\infty} P_{+1}^{\infty , \ell m \omega_{mn}}(r) , & r \ge \rmax , \\
 \alpha_{+1}^{h} P_{+1}^{h , \ell m \omega_{mn}}(r) , & r \le \rmin , 
\end{cases}
\eeq
where $\alpha_{+1}^{\infty} = 2 \pi \widetilde{\mathfrak{C}}_n$ is the coefficient stated in Eq.~(\ref{eq:alpha-inf}), and the coefficient $\alpha_{+1}^{h}$ is also derived via the argument above, \emph{mutandis mutatis}. 

Repeating the steps above for spin-weight $-1$, it is straightforward to derive an expression for the Maxwell scalar $\phi_2$ ,
\beq
2 (r-i a \cos \theta)^2 \phi_2 = \Delta^{-1} \sum_{\ell m n} S_{-1}^{\ell m \gamma}(\theta) e^{- i \omega_{mn} t + i m \phi}  
\begin{cases}
 \alpha_{-1}^{\infty} P_{-1}^{\infty , \ell m \omega_{mn}}(r) , & r \ge \rmax , \\
 \alpha_{-1}^{h} P_{-1}^{h , \ell m \omega_{mn}}(r) , & r \le \rmin . 
\end{cases}
\eeq
From these asymptotic expressions for the Maxwell scalars, and using the standard unit normalisation of the homogeneous Teukolsky functions $P_{\pm1}^{h/\infty}(r)$, we can immediately obtain the fluxes in Eqs.~(\ref{eq:fluxes}) by using the standard derivations in Refs.~\cite{Press:1973zz,Teukolsky:1973ha,Teukolsky:1974yv} (see also Appendix in Ref.~\cite{Torres:2020fye}).

\section{Energy, angular momentum and orbital parameters\label{appendix:ELpe}}
Here we list the relationship between $(E,L)$ and $(p,e)$, as described in Ref.~\cite{Warburton:2011hp}. The energy $E$ and angular momentum $L$ of a particle in an equatorial orbit, locally tangent to a geodesic, are
\begin{align}
    E &=\left[1-\left(\frac{M}{p}\right)\left(1-e^2\right)\left\{1-\frac{x^2}{p^2}\left(1-e^2\right)\right\}\right]^{1 / 2} \label{Eq:Eep}\\
    L &= x + a E\label{Eq:Lep}
\end{align}
 where $x= x(a,p,e)$ is the rather complicated function given as:
\begin{equation}
    x=\left[\frac{-N-\operatorname{sign}(a) \sqrt{N^2-4 F C}}{2 F}\right]^{1 / 2},
\end{equation}
with
\begin{align}
F(p, e)&=\frac{1}{p^3}\left[p^3-2 M\left(3+e^2\right) p^2+M^2\left(3+e^2\right)^2 p\right. 
\left.-4 M a^2\left(1-e^2\right)^2\right] \\
N(p, e)&=\frac{2}{p}\left\{-M p^2+\left[M^2\left(3+e^2\right)-a^2\right] p\right. 
\left.-M a^2\left(1+3 e^2\right)\right\} \\
C(p)&=\left(a^2-M p\right)^2
\end{align}

\section{Orbits remaining until plunge\label{appendix:orbits}}
A useful quantity to calculate is the number of orbits remaining until plunge. To find the number of orbits remaining till plunge one must first calculate the total number of orbits completed. This is done by evaluating $\phi(\tau)/2\pi$ at $\tau = \tau_f$, the proper time at which the the inspiral hits the separatrix. $\phi(\tau)$ is calculated by integrating its derivative between 0 and $\tau$. For geodesics in the equatorial plane one can write $\dot{\phi}$ as a function of the constants of motion \cite{chandrasekhar1998mathematical}:
\begin{equation}
    \dot{\phi} = \frac{1}{\Delta} \left(\frac{2aM}{r}E+L \left(1-\frac{2 M}{r}\right)\right)
\end{equation}
Now, for equatorial orbits we have a relationship between $r$ and $(p,e)$ given by Eq.~\eqref{eqn:r(e,p)}. Substituting this in and expanding this as a series in $\cos(\chi)$ one gets the following approximation:
\begin{equation}
    \dot{\phi} \approx \frac{2 a E+L (p-2)}{p \left(a^2+M^2 (p-2) p\right)} + \mathcal{O}(\cos{\chi})
\end{equation}
Note, here $e, p, E, L$ are all functions of $\tau$. $E$ and $L$ are both found through the \verb|KerrGeodesics|\cite{niels_warburton_2023_8108265} package, in terms of the functions $e(\tau)$ and $p(\tau)$, which are themselves evaluated on the trajectory (see Section~\ref{sec:InspiralModel}). Thus we have that:
\begin{equation}
    \phi(\tau_0) \approx \int\limits_0^{\tau_0} \frac{2 a E(\tau)+L(\tau) (p(\tau)-2)}{p(\tau) \left(a^2+M^2 (p(\tau)-2) p(\tau)\right)} \dd \tau.
\end{equation}
We define $n_f = \phi(\tau_f)/2\pi$. To find the point on trajectory with $n$ orbits remaining till plunge one solves $\phi(\tau)/2\pi = n_f -n$ for $\tau$ using a root finding algorithm such as Mathematica's \verb|FindRoot[]|.  We expect the number of orbits remaining to scale in inverse proportion to the square of the charge-to-mass ratio $q/m$.

\bibliographystyle{apsrev4-1}
\bibliography{kerrEMadiabatic}
\end{document}